\documentclass[reprint,amsmath,amssymb,aps,prl]{revtex4-1}
\usepackage{graphicx}
\usepackage{bm}

\usepackage[british]{babel}

\begin{document}

\preprint{APS/123-QED}

\title{Quasi-ballistic transport of Dirac fermions in a Bi$_2$Se$_3$ nanowire}
\date{\today}

\author{J. Dufouleur$^1$, L. Veyrat$^1$, A. Teichgr\"aber$^1$, S. Neuhaus$^1$, C. Nowka$^1$, S. Hampel$^1$, J. Cayssol$^{2 3}$, J. Schumann$^1$, B. Eichler$^1$, O. Schmidt$^1$, B. B\"uchner$^1$ and R. Giraud$^{1 4}$}
\affiliation{
$^{1}$Leibniz Institute for Solid State and Materials Research, IFW Dresden, 01171 Dresden, Germany\\
$^{2}$Max-Planck-Institut f\"ur Physik Komplexer Systeme, N\"othnitzer Str. 38, 01187 Dresden, Germany\\
$^{3}$LOMA, University Bordeaux 1, F-33045 Talence, France\\
$^{4}$CNRS - Laboratoire de Photonique et de Nanostructures, Route de Nozay, 91460 Marcoussis, France\\}


\begin{abstract}
Quantum coherent transport of Dirac fermions in a mesoscopic nanowire of the 3D topological insulator Bi$_2$Se$_3$ is studied in the weak-disorder limit. At very low temperatures, many harmonics are evidenced in the Fourier transform of Aharonov-Bohm oscillations, revealing the long phase-coherence length of surface states. Remarkably, from their exponential temperature dependence, we infer an unusual 1/$T$ power law for the phase coherence length $L_\varphi(T)$. This decoherence is typical for quasi-ballistic fermions weakly coupled to the dynamics of their environment.
\end{abstract}


\maketitle

In a mesoscopic conductor, quantum corrections to the classical conductance arise from the phase-coherent transport of delocalized carriers \cite{Altshuler1985,Akkermans2007}. Their amplitude and temperature dependence reveal important informations on both the phase coherence length $L_{\varphi}$ and the dimensionality of quantum coherent transport. Importantly, the temperature dependence of $L_{\varphi}$ is determined by the dominant mechanism responsible for decoherence. For massive quasi-particles, quantum coherence is usually limited by electron-electron interactions at very low temperatures, which result in a power-law temperature dependence as $L_{\varphi} \propto T^{-\alpha}$. In the diffusive regime and for quasi-1D coherent transport, $\alpha=1/3$ for a wire geometry \cite{Altshuler1982} or $\alpha=1/2$ for a ring shape \cite{Ludwig2004,Texier2005,Ferrier2008}. However, in the rare case of ballistic transport of fermions with a weak coupling to the environment, such as for quasi-1D ballistic rings \cite{Hansen2001} or for edge states in the Integer Quantum Hall regime \cite{Roulleau2008}, decoherence is dominated by the fluctuations of the environment. In this regime, $L_{\varphi}$ follows a $1/T$ dependence \cite{Seelig2001}. For massless Dirac fermions, as found in graphene or in 3D topological insulators, the ratio between the interaction energy to the environment and the kinetic energy is decreased due to the relativistic nature of free carriers. Besides, the coupling strength is further weakened in a nanostructure, when quantum confinement reduces the carrier density of states. Yet, little is known on decoherence at very low temperatures, and the weak coupling regime of Dirac fermions to their environment was not evidenced to date. In graphene nanoribbons, this is particularly difficult to achieve, due to both the band gap opening in the quasi-1D limit and to charge inhomogeneities \cite{DasSarma2011}. In a 3D topological insulator however, surface states Dirac fermions are not altered by perturbations which do not break the time-reversal symmetry \cite{Moore2010,Hasan2010a,Qi2011}, and the spin chirality of the surface states favors an even weaker coupling to the environment. Moreover, nanostructures with a good crystalline quality can be grown in a bottom-up approach, and are therefore well suited to study quantum coherent transport in the weak-disorder limit.

Electronic surface states in a 3D topological insulator (TI) are 2D spin-chiral Dirac fermions, which were first predicted theoretically \cite{Fu2007,Moore2007,Roy2009,PhysRevB.76.045302}, and then convincingly evidenced in the strong topological insulators Bi$_2$Se$_3$ and Bi$_2$Te$_3$, by means of both angle-resolved photo-emission spectroscopy \cite{Xia2009,Hsieh2009,Chen2009} and electrical transport measurements \cite{Cheng2010,Analytis2010,Checkelsky2011,Steinberg2010,Cho2011,Sacepe2011}. In the latter case, the electrical detection of surface states requires to measure nanostructures, so to reduce the residual bulk states contribution to the conductance. Remarkable transport properties of surface Dirac fermions were thus revealed, such as the ambipolar nature of charge carriers \cite{Cheng2010,Steinberg2010,Cho2011,Sacepe2011} and their relativistic nature \cite{Cheng2010,Sacepe2011}. Quantum coherent transport further provided a direct evidence of the existence of surface states in nanostructures of a 3D TI, through the observation of well-defined Aharonov-Bohm oscillations \cite{Peng2010,Xiu2011,Bardarson2012}. Nevertheless, all these previous studies were performed only in the relatively-high temperature regime ($T>1$~K), where electron-phonon inelastic processes dominate. This hinders the intrinsic mechanism that ultimately determines the quantum transport properties of spin-chiral Dirac fermions in nanostructures of 3D topological insulators. Besides, despite interesting studies at very low temperatures of the weak-antilocalization in the diffusive transport regime of thin films \cite{Wang2011} or flakes \cite{Checkelsky2011,Steinberg2011} of Bi$_2$Se$_3$, the issue of decoherence was not addressed yet.

In this letter, we report on a detailed analysis of quantum interferences effects on the conductance of a weakly-disordered single-crystalline Bi$_2$Se$_3$ nanowire, measured in a finite magnetic field and at very low temperatures. A strong contribution from Aharonov-Bohm oscillations, which only results from surface Dirac fermions, superimposes on universal condutance fluctuations (UCF). Using a 3D-vector magnetic field, we reveal a dimensionality effect which is the signature of the quasi-1D nature of quantum coherent transport, showing that $L_{\varphi}$ is longer than the transverse dimension of the nanowire over the broad temperature range studied. This very long phase coherence length ($L_{\varphi}>2\mu$m at $T=50$~mK) gives rise to many harmonics in the Aharonov-Bohm oscillations, as revealed by a Fourier transform analysis. Most important, the intrinsic exponential decay of their amplitudes with temperature is observed over two orders of magnitude, from 20~mK up to about 2~K. Remarkably, this allows us to clearly evidence the 1/$T$ dependence of the phase coherence length, which is typical of the quasi-ballistic transport of fermions that are weakly coupled to their environment.

Nanostructures of Bi$_2$Se$_3$ were prepared by Chemical Vapor Deposition (CVD), in sealed quartz tubes, directly onto Si/SiO$_x$ substrates and without any catalyst. The CVD growth conditions on amorphous SiO$_x$ were optimized to obtain ultra-thin (thickness $d\geq 6$~nm) and/or narrow (width $w\geq 50$~nm) single crystals, which can extend over tens of microns on the SiO$_x$ surface. Standard e-beam lithography and lift-off process were used to contact an individual nanostructure, and then to measure its transport properties. From Hall measurements on thin flakes, we found a typical $n$-type doping of about 10$^{19}$~cm$^{-3}$, that reveals a smaller Se-vacancy defects density as compared to values previously reported for thin flakes exfoliated from macroscopic single crystals. Despite the reduced bulk states density, the Fermi energy remains close to the bottom of the conduction band, so that 3D states can also contribute to charge transport in our nanostructures. However, the study of quantum coherent transport is simplified for a nanowire or narrow nanoribbon geometry if the magnetic field is applied parallel to its length. In this case, the surface states and bulk states contributions to quantum interferences are separable, as discussed below, which allows us to study the quantum coherence of 2D Dirac fermions only.


\begin{figure}[!t]
\includegraphics[width=1\columnwidth]{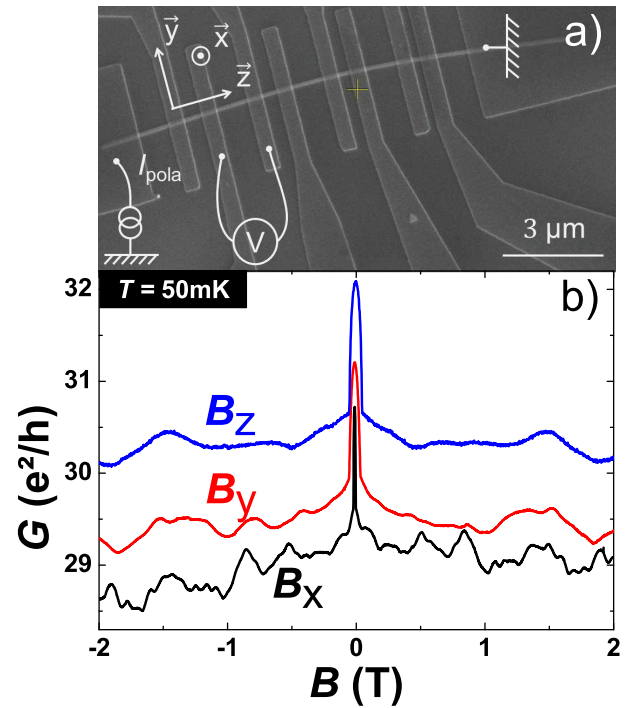}
\caption{\textbf{a}, SEM picture of the nanowire, with a rectangular cross-section ($w=90$~nm, $d=50$~nm) and a length of about 18~$\mu$m. \textbf{b}, Magneto-conductance measured along the $\vec{x}$, $\vec{y}$ or $\vec{z}$ axes of the 3D magnet at $T=50$~mK. The measurements in $B_x$ and $B_z$ are shifted, for clarity.}
\label{fig1}
\end{figure}

In this work, we studied the quantum transport properties of such a Bi$_2$Se$_3$ nanowire (width $w=90$~nm, thickness $d=50$~nm), as shown in Fig.~\ref{fig1}a), the dimensions of which were measured by scanning electron microscopy (SEM) and atomic force microscopy. Good ohmic Ti($10$~nm)/Al($100$~nm) contacts were obtained after a surface deoxidation (wet etching). Quantum transport properties were measured in a four-probe geometry with lock-in amplifiers, from 4.2~K down to the 20~mK base temperature of a $^3$He/$^4$He dilution refridgerator, using a small enough current polarisation to avoid electronic heating. Most important for our study, the magneto-conductance was investigated in all three relevant orientations of the applied magnetic field (along the nanowire axis, $B_z$, or perpendicular to it, $B_{x,y}$), by using a 3D 2T-vector magnet (yet with a main induction of 6T along the $\vec{z}$ direction). 

Below about 1~K (corresponding to the onset of superconductivity in the Al contacts), the high transparency of the ohmic contacts is directly revealed by the excess conductance at zero magnetic field due to a superconducting proximity effect. As seen in Fig.~\ref{fig1}b) at $T=50$~mK, for a conductor length of 840~nm, this excess conductance rapidly drops when a magnetic induction $B$ is applied, and disappears beyond the critical field of TiAl nanocontacts (which is enhanced by a reduced dimensionality effect if measured along the $\vec{y}$ or $\vec{z}$ directions). Importantly for quantum transport measurements, the low resistance $R_C$ of ohmic contacts is not negligible as compared to the resistance of the nanowire. This results in a strong influence of $R_C$ on the root-mean-square amplitude of the measured quantum corrections to the conductance $\delta G_{meas}$. This is known for two-probe measurements, but it is also valid for four-probe measurements in the case of a partial flow of the current through metallic contacts. In this case, the conductance is symmetric in $B$, and the  absolute amplitude of quantum corrections $\delta G_{QC}$ to the nanowire conductance $G_{NW}$ is renormalized to a reduced value $\delta G_{meas}=\frac{\delta G_{QC}}{(1+R_C \times G_{NW})^2}$. For the measure of Fig.~\ref{fig1}b), given that $\delta G_{QC}\approx G_0=e^2/h$ (see below), we find $R_C \approx 0.83/G_{NW} \approx 277$~$\Omega$, a value which is in good agreement with the one infered from the conductance itself. Note that this scaling factor is nearly temperature independent, so that it does not influence the temperature dependence of $\delta G_{meas}$.

Beyond the sharp conductance peak around zero field, large and reproducible conductance fluctuations can be seen in Fig.~\ref{fig1}b), for all field directions. At small fields, a weak-antilocalization (WAL) correction to the classical conductance is observed. At higher fields, the WAL is destroyed and only two other quantum interference effects remain: the aperiodic UCF and, if the field is applied along the nanowire axis ($\vec{z}$ direction), the periodic Aharonov-Bohm (AB) oscillations. Let us first discuss the dimensionality effects observed on the UCF. As seen in Fig.~\ref{fig1}b), the correlation field $B_C$ of aperiodic conductance fluctuations depends on the orientation of the applied field. This is a direct signature of the quasi-1D nature of quantum coherent transport. Indeed, $B_C$ can be related to the largest coherent loops perpendicular to the applied field. If $L_{\varphi}\ge$($w$,$d$), the condition for quasi-1D coherent transport, then $B_C \approx \phi_0/S$ is determined by $S=L_{\varphi}\times w$, $L_{\varphi}\times d$ or $w\times d$, for a field applied along $\vec{x}$, $\vec{y}$ or $\vec{z}$, respectively. $\phi_0=h/e$ is the flux quantum. This is evidenced in Fig.~\ref{fig1}b) and gives a direct signature of the long phase coherence length. As explained above, the amplitude of the quantum fluctuations strongly depends on the contact resistance, and it can vary for different pairs of contacts separated by a similar length. At $T=50$~mK and for large-impedance ohmic contacts separated by a micron, the amplitude of quantum fluctuations can be as large as the quantum of conductance $G_0=e^2/h$, which indeed confirms that $L_{\varphi}$ exceeds the micron scale at very low temperatures. Note that the quasi-1D coherent regime is observed in the entire temperature range studied, which therefore sets a lower bound $L_{\varphi}\ge 50$~nm at $T=4.2$~K \footnote{This is in agreement with previous evaluations of $L_{\varphi}$ in the high-temperature regime \cite{Peng2010,Wang2011,Steinberg2011} \label{Note1}}. Given this value and from the temperature dependence of $L_{\varphi}$ (see below), we find a lower bound $L_{\varphi}\ge2$~$\mu$m at $T=50$~mK.

\begin{figure}[!t]
\includegraphics[width=1\columnwidth]{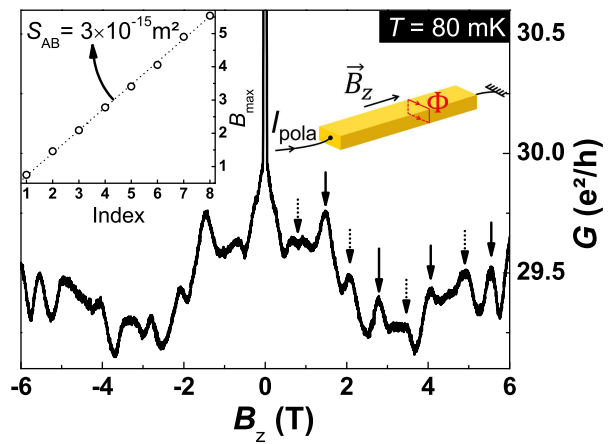}
\caption{Magneto-conductance $G$($B_z$), $B_z$ being parallel to the nanowire. Dominant periodic oscillations in $\phi_0$ are clearly observed (full arrows) and smaller ones in $\phi_0/2$ are also visible (dotted arrows). The inset shows the field position of all periodic maxima of these fundamental and first harmonics in the AB oscillations. The linear slope corresponds to $(\phi_0/2)/S_{AB}$, where $S_{AB}$ is the effective electrical cross-section enclosed by surface states.}
\label{fig2}
\end{figure}

If the magnetic field is applied along the nanowire axis ($\vec{z}$ direction), the magneto-conductance reveals the dominant contribution of AB oscillations. Importantly, these arise from surface states only, and coexist with UCF which only result from bulk states. This gives a unique way to study the quantum coherent transport properties of 2D Dirac fermions, independently from bulk-state transport. Clear $\phi_0$-periodic AB oscillations are indeed observed in Fig.~\ref{fig2}, and the inset points at the regular position of the main maxima corresponding to the fundamental ($n=1$) and first ($n=2$) harmonics. 
The slope of the linear fit corresponds to an electrical cross-section $S_{AB}=3\times10^{-15}$m$^{-2}$ which is somewhat smaller than the measured cross-section $S=4.5 \times10^{-15}$m$^{-2}$ of the nanowire. This suggests that the interface states are located about 5~nm below the surface, which is a reasonable assumption considering both the surface oxidation and their evanescent behavior into the bulk of Bi$_2$Se$_3$. 
As reported earlier at higher temperatures, this gives a direct evidence of the existence of the surface states, which set a well-defined magnetic flux enclosed by coherent paths, contrary to bulk states trajectories. Interestingly, the measure of the AB oscillations at very low temperature further reveals new properties of these surface states.

\begin{figure}[!t]
\includegraphics[width=1\columnwidth]{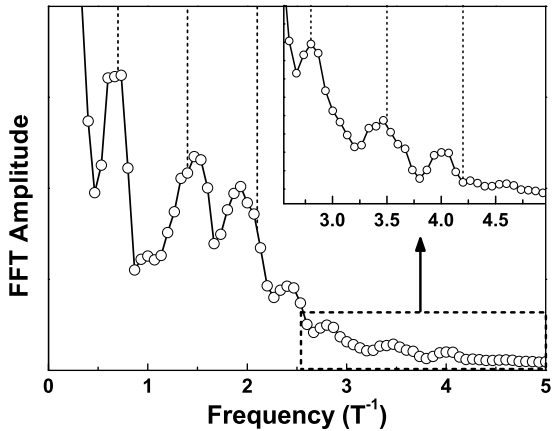}
\caption{Fast-Fourier Transform of $G$($B_z$). Vertical dashed lines indicate the AB harmonics from the fundamental ($n=1$) to the highest harmonic observed ($n=6$). The non-monotonic UCF contribution is reduced by an averaging process, as described in the text.}
\label{fig3}
\end{figure}

In a nanowire, UCF and AB oscillations are not directly separable in the measure of $G(B_z)$, because the correlation field of aperiodic fluctuations is comparable to the AB period. However, the Fourier transform analysis of $G(B_z)$ measured at $T=50$~mK allows us to evidence the many harmonics in the AB oscillations which, in a semi-classical approach, result from the multiple coherent loops that can be experienced by the surface states when propagating along the perimeter of the nanowire. These regular peaks in Fig.~\ref{fig3} emerge from a non-monotonous background induced by the UCF, because the magnetic-field range studied is finite. In order to reduce the UCF contribution to the Fourier spectrum and to quantitatively extract the information on the AB fluctuations only, we calculated the Fourier transform over a partial field range only, within a field window which is then shifted by $\delta B \leq B_C$ to repeat the Fourier analysis, and we finally averaged all the spectra. Besides, we used a so-called flat-top window to reduce the frequency spread and to efficiently smooth out non-periodic contributions in $G(B_z)$. Note that AB harmonics are independent from the procedure. This allow us to clearly separate the AB harmonics in the Fourier transform and to quantitatively study their position and temperature dependence.

\begin{figure}[!t]
\includegraphics[width=1\columnwidth]{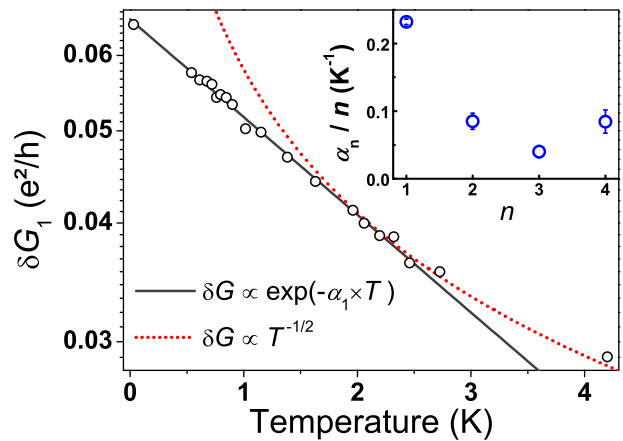}
\caption{Temperature dependence of the integrated fundamental harmonic, showing a clear exponential decay from 30~mK up to about 2.5~K (open dots: measurements; line: $A \times \exp {-\alpha_1 T}$) fit, excluding the $T=4.2$~K data point). The high-temperature $1/\sqrt{T}$ regime is shown as a dotted line. Higher harmonics follow a similar behavior with a different slope $\alpha_n$, which does not scale with $n$ (see inset).}
\label{fig4}
\end{figure}

A dominant peak is evidenced at about 0.7~T$^{-1}$ in Fig.~\ref{fig3}, which corresponds to a $\phi_0$-periodic signal for an electrical cross-section of about 3~$\times 10^{-15}$~m$^2$. This periodicity of AB oscillations is a direct signature of the coherent transport in the weak disorder limit \cite{Bardarson2010} \footnote{The influence of both disorder and the Fermi energy position on the $\phi_0$ periodicity of AB oscillations in 3D TIs was discussed in \cite{Bardarson2010,Zhang2010a}. In our study, the Fermi energy position is far away from the Dirac degeneracy point. In this case, it was shown that the $\phi_0$ periodicity is a direct signature of the weak disorder limit \cite{Bardarson2010} \label{Note2}}. At higher frequencies, the other peaks observed correspond mostly to harmonics of the fundamental ($n=1$), which can be labelled from $n=2$ to $n=6$. Such AB harmonics were found in all five different sets of contacts measured. The evolution of the amplitude of the harmonics is typical of AB oscillations, but the exact peak positions can be slightly shifted from a simple linear evolution with $n$ (suggested in Fig.~\ref{fig3} by vertical dashed lines). Although we cannot totally rule out a remaining influence of the UCF background, these shifts seem to be an intrinsic property of the harmonics, the understanding of which requires further investigations. 
From the temperature dependence of the AB oscillations, we can study the decoherence of the 2D spin-chiral Dirac fermions \footnote{Note that thermal smearing does not influence the temperature dependence of $\delta G_{rms}$ in the temperature range studied. A lower bound $T_S$ for the temperature below which this effect is negligible is given by $\frac{\hbar v_F}{2(w+d)}$. In Bi$_2$Se$_3$, this gives $T_S\ge10$~K for $v_F \approx 5\times10^5$~m$\cdot$s$^{-1}$ \cite{Analytis2010} \label{Note3}}. First, we evidence a departure at very low temperatures from the previously reported $1/\sqrt{T}$ power law and demonstrate the intrinsic exponential behavior below about 2.5~K down to our 20~mK base temperature, that is, over two orders of magnitude. This is shown in Fig.~\ref{fig4} for the fundamental harmonic ($n=1$). A similar behavior is found for all harmonics (with different slopes $-\alpha_n$), as well as for other pairs of contacts with slightly different contact resistances. Interestingly, the inset in Fig.~\ref{fig4} shows that the ratio $\alpha_n/n$ is not a constant, contrary to what is known for the diffusive regime \cite{Akkermans2007} and also observed with quasi-1D ballistic rings \cite{Hansen2001}. 

Most importantly, the e$^{-\alpha_n T}$ dependence is directly related to an unusual temperature dependence of the phase coherence length, with $L_{\varphi} \propto 1/T$. This reveals that decoherence is limited by a weak coupling to a fluctuating environment \cite{Seelig2001}, as can be expected for spin-chiral Dirac fermions with a limited phase space density. To our knowledge, such a seldom situation was previously reported for massive quasi-particles, but found only in the case of quasi-1D ballistic rings with a limited number of transverse modes \cite{Hansen2001} and for edge states in the Integer Quantum Hall regime \cite{Roulleau2008}. Therefore, our observation of such an intrinsic limit to the decoherence of surface states at very low temperatures gives not only a strong proof of the quasi-ballistic transport of Dirac fermions in a CVD-grown Bi$_2$Se$_3$ nanowire, but it also suggests the generality of the $1/T$ dependence of $L_{\varphi}$ in the multi-channel quasi-ballistic regime (in the nanowire studied here, we estimate the number of transverse modes to be of about 30). Given that $L_{\varphi}=v_F\tau_{\varphi}$ for ballistic transport, such an unusual temperature dependence results from a decoherence mechanism which gives a temperature dependence of the dephasing time $\tau_{\varphi}\propto 1/T$, as can be induced by the dynamics of scattering centers \cite{Imry1999}. 
Another important outcome is that the evidence of quasi-ballistic transport of spin-chiral Dirac fermions over the perimeter of the Bi$_2$Se$_3$ nanowire ($\approx$240~nm), a much longer lengthscale compared to the inter-defect distance (of about 5~nm for a residual doping of 10$^{19}$~cm$^{-3}$), is a direct signature of the very weak scattering potential on Se-vacancy defects. The strength of disorder being the consequence of both the disorder density and the scattering potential, this shows that the weak-disorder limit is actually achieved because of the weak scattering of surface states in 3D topological insulators.

Such findings confirm the interest of spin-chiral 2D Dirac fermions in a 3D topological insulator to further investigate quantum coherent transport in new regimes otherwise not reachable.


\begin{acknowledgments}
We thank Dr. Ingolf M\"onch for technical support. 
J.~C. acknowledges the support from EU/FP7 (contract TEMSSOC) and from ANR (project No. 2010-BLANC-041902, ISOTOP).
\end{acknowledgments}



%

\end{document}